\documentclass[twocolumn]{article}

\usepackage{amssymb}
\usepackage{amsmath}
\usepackage{epsfig}

\begin{document}
\title{Bethe-Peierls Approximation for Linear Monodisperse
Polymers Re-examined}
\author{F. F. Semeriyanov and G. Heinrich \\
Leibniz Institute of Polymer Research Dresden, Hohe Str. 6,
01069 Dresden, Germany}
\date{\today}

\maketitle
\begin{abstract}
Bethe-Peierls approximation, as it applies to the thermodynamics of
polymer melts, is reviewed. We compare the computed configurational entropy of
monodisperse linear polymer melt with Monte Carlo data available in
literature. An estimation of the configurational
contribution to the total liquid's Cp is presented. We also discuss
the relation between Kauzmann paradox and polymer semiflexibility.
\end{abstract}
\section{Introduction}

In 1948, Kauzmann \cite{Kauzmann1948} recognized a peculiar fact
about the thermodynamic behavior of liquids in the vicinity of glass
transition. The configurational entropy of supercooled liquids drops very
rapidly with temperature decrease and this leads to the entropy crisis. It looks like the
entropy would drop to zero at $T>0$ and then
something should happen to the system in order to avoid the entropy
becoming negative. The problem has attracted much theoretical
attention after the work by Gibbs and Di Marzio
\cite{Gibbs-DiMarzio}. They demonstrated, by means of a statistical
calculation, that the entropy of a disordered polymer liquid
extrapolates to negative values at low temperatures. This entropy crisis
violates the Nernst postulate and makes the problem very puzzling.
To resolve this problem, they proposed that the second-order
transition to a unique state, the ideal glass, would occur at the
temperature where the configurational entropy turns to zero. Gujrati
and Goldstein \cite{Gujrati-Goldstein1,Gujrati-Goldstein2} criticized the Gibbs-DiMarzio
theory for the absence of a crystalline state. The
crystal to liquid states and their metastable extensions in a polymer system
were subsequently captured by Corsi and Gujrati \cite{CorsiGjrati1,CorsiGjrati2} with
a demonstration of the configurational entropy crisis
at low temperatures for the metastable liquid state. They pointed out that
the unusual ideal-glass transition in the metastable region does not violate
the thermodynamic laws provided that the stable crystalline state exists below
the melting temperature, $T_{M}>T_{K}$, see also \cite{GujratiCondMat}. In addition, developed by Freed and co-workers
\cite{FreedEtAl} the lattice cluster theory predicts that the configurational entropy would extrapolate to zero at a positive
temperature \cite{DudowiczEtAl}, as in the Gibbs-DiMarzio theory.

The present work concerns only the equilibrium melt of monodisperse polymers, whereas the equilibrium polymerization requires additional parameters, e.g. polymer polydispersity. Monodisperse polymers can be prepared in laboratory and, more importantly, can be simulated using Monte Carlo methods. In the present work, the crystal is not captured, even though it is possible to construct such a state for monodisperse polymers. Strictly speaking, the total configurational entropy is the sum over the entire energy spectrum. However, crystal, glass and liquid states are usually pictured to occupy separate basins of the energy landscape \cite{Goldstein}, which suggests that the corresponding systems can be treated separately. It is important to understand whether or not matter could exist in a form of a deeply supercooled liquid, even though the configurational entropy of glass may differ form that of the liquid at the same temperature \cite{GujratiRecent1,GujratiRecent2,JohariRecent}.

In an attempt to address to this problem, an analytic calculation of configurational
entropy of a polymer melt is presented in order to provide a framework for a more systematic
study. To solve the problem we use the Bethe-Peierls approximation,
which is a lattice gas theory neglecting all closed loops of the regular lattice. More precisely, we want
to investigate the interplay of the effects due to closed loops of a regular lattice, polymer
chain length and chain semiflexibility, and to relate this to the entropy crisis.

\section{Model}

We consider the semiflexible linear polymers of the same length $x$ distributed on the lattice.
The penalty for a chain bend is introduced by means of the Flory's
flexibility \cite{Flory1956,Flory1982}, $f$. The
interactions among monomers are taken into account by assigning the energy, $V<0$ (attractive
interaction), to the nearest-neighbor pairs of occupied sites. The
Bethe-Peierls approximation \cite{BethePeierls1,BethePeierls2} was originally
developed for single site species. Later, Chang \cite{Chang1939}
applied the method to diatomic molecules. The aim of the present
paper is to extend the Chang's method to polyatomic species.
The entropy expression is found to
reduce to the Flory-Huggins form \cite{FloryHuggins1,FloryHuggins2} in the
appropriate limit: $V\rightarrow0$ and $z\rightarrow\infty$, provided that $zV$=const, where $z$ is the
lattice coordination number.

\section{Bethe-Peierls approximation}

The main idea of the method is to calculate the grand partition
function (GPF) for configurations of an aggregate composed of a
molecule occupying a lattice site (internal site) and its $z$
neighbors (surface sites) in the framework of a mean-field theory:
An effective field is superimposed on the surface molecules by
molecules outside the aggregate. It is determined making use of the
condition of equal probability for the internal molecule as well as
for any molecule from the aggregate surface to be in a particular
configuration. After determination of GPF other quantities such as
energy (enthalpy at constant pressure), specific heat, etc. follow
at once.

The effective interaction parameter, $V$, is defined as the energy
of interaction between two occupied sites that are nearest-neighbors
to each other irrespective of whether they are connected by a bond
or not. To account for presence of free volume, the lattice-hole
model is employed in which both monomers and holes (empty lattice
sites) populate a lattice with the occupancy controlled by the
system chemical potential, $\mu$. The polymer semiflexibility is
incorporated through $U_{f}$ being the energetic penalty for
creation of a chain bend. Based on this information, a general form
of GPF can be given:
\begin{equation}
Z=\sum \Omega (n,X,N_{g},N)\eta ^{X}\lambda ^{n}w_{f}^{N_{g}},
\label{GPF}
\end{equation}
where $\eta =\exp (-V/k_{\text{B}}T)$ is the Boltzmann weight for
the interactions between occupied sites, $\lambda =\exp (\mu%
/k_{\text{B}}T)$ is the absolute activity, $w_{f}=\exp (-U_{f}
/k_{\text{B}}T)$ is the Boltzmann weight for the flex energy, $n$ is
the total number of occupied sites, $X$ is the number of
nearest-neighbor pairs of occupied sites, $N_{g}$ is the total
number of the gauche configurations, $\Omega $ is the number of
arrangements on a lattice of $N$ sites that have the same numbers of
$n$, $X$, and $N_{g}$.

For comparison purposes, it is instructive to begin by presenting
the original Bethe-Peierls technique. Next, we are going to present
a modification due to the chain connectivity. At first, the
aggregate is pictured as a site surrounded by $z$ surface sites. A
set of numbers is defined: $\{\theta _{i}\}=\theta _{0},\theta
_{1},...,\theta _{z}$, which are 0 or 1 according as the
corresponding sites are empty or occupied, where the index $i=0$ is
reserved for the internal site. A configuration with the internal
site being unoccupied $(0;\theta _{1},...,\theta _{z})$ contributes
to GPF as the sum over all arrangements on the surface sites
\begin{equation}
    {\sum
}_{\{\theta_{i}\}} \lambda ^{\theta _{1}+...+\theta _{z}}\psi
(\theta _{1},...,\theta _{z}),
\end{equation}
where $\psi$ is the field that
couples the aggregate configurations to the external distribution.
Presuming that there are no sites occupied by the segments of the
same chain among the surface sites, two types of correlations
generated by a local configuration can be identified: the occupancy
of a surface site 1) means the possibility of the bond presence
between the given site and an external nearest-neighbor site, 2)
induces the mer-mer interaction with the occupied external sites. An
empty surface site does not induce any correlations to the exterior.
These facts can be generally taken into account presuming that
$\psi$ is given by
\begin{equation}
    \psi =c\xi ^{\theta _{1}+...+\theta _{z}},
\end{equation}
where $c$ is some constant. From this follows that for the
configuration $(0;\theta_{1},...,\theta _{z})$ GPF is given by
\begin{equation}
\underset{\{\theta_{i}\}} {\sum } c(\lambda \xi )^{\theta
_{1}+...+\theta _{z}}=c(1+\lambda \xi )^{z}. \label{GPF_sm_cluster}
\end{equation}

To study the linear chains, we consider the aggregate composed of
$x$ sites and their $z^{\prime}=2(z-1)+(x-2)(z-2)$ next neighbors.
The modification was inspired by the papers published quite some
time ago \cite{Chang1939,Miller1,Miller2}.

The GPF for the configuration $(1,...,1;\theta_{x+1},...,\theta
_{x+z^{\prime}})$ can be represented in the following general form:
\begin{equation}
\underset{i_{2}i_{3}...i_{x}} {\sum} \underset{\{{\sum} ^{j}\}}
{\sum} c_{x}\lambda^{x} \eta ^{x-1}(\lambda\eta\xi_{x})^
{{\sum}^{2}\theta + ... +{\sum}^{x}\theta}, \label{GPF_to_calc}
\end{equation}
where the first sum is over configurations permitted by the linear
chain topology, the second sum is over configurations on the surface
sites, the factor $\eta ^{x-1}$ takes into account the interactions
among the segments of the chain inside the aggregate, and
\begin{equation}
    {\sum}^{j}\theta=\theta_{1}+..+\theta_{j-1}+\theta_{j+1}+\theta_{x}.
\end{equation}
In order to calculate (\ref{GPF_to_calc}), we propose a recursive
method analogous to the Cayley tree technique
\cite{Ryu-Gujrati1997}. Accordingly, the aggregate composed of a
site and its $z$ neighboring sites corresponds to the center of the
Cayley tree. As before, the configurations with the central site
being unoccupied are included in GPF by the term $c(1+\lambda \xi
)^{z}$, where $\xi$ represents the effective field for the case when
no bond is present inside the aggregate. For the case when the
central site is occupied by an end-point or by a middle group, we
define $\chi _{k}$ to account for correlations induced on the
aggregate due connectivity of the chain; where $k$ is the number of
bonds in the chain piece extending beyond the surface site of the
aggregate. The GPF can be formulated as follows:
\begin{eqnarray}
Z &=&c(1+\lambda \xi )^{z}+cz(1+\lambda \eta \xi )^{z-1}\lambda
^{2}\eta
\chi _{x-1}  \notag \\
&&+c\frac{z}{2}r_{f}(1+\lambda \eta \xi )^{z-2}\lambda ^{3}\eta
^{2}\underset{k=1}{ \overset{x-2}{\sum }}\chi _{k}\chi _{x-k-2},
\label{rec GPF}
\end{eqnarray}
where the factor $z$ in the second term is the number of
orientations for the bond extending from the end-point occupying the
internal site of the aggregate; in the third term, the factor
\begin{equation}
r_{f}=w_{f}(z-2)+1 \label{r_f}
\end{equation}
takes into account the semiflexibility. The latter is derived making
use of the assumption that the number of trans configurations in the
aggregate is equal to $z/2$. In the case when $z$ is an even number
and the surface sites, situated around the internal (central) site,
are equally spaced, $z/2$ is easily identified as the number
configurations of two bonds with the 180 degree angle between them.
Note that this simplified picture does not apply when $z$ is an odd
number since a construction of trans configurations with the 180
degree angle between the bonds is not possible for this case.
Proceeding further with our description of (\ref{rec GPF}), we
justify the statistical weight $zr_{f}/2$ by means of the relation
\begin{equation}
\left[ \frac{z(z-1)}{2}-\frac{z}{2}\right]
w_{f}+\frac{z}{2}=\frac{z}{2}r_{f},
\end{equation}
where $z(z-1)/2$ is the total number of configurations for two bonds
inside the aggregate. In order to evaluate GPF, we obtain the Behte
lattice recursive relations for $\chi _{k}$:
\begin{eqnarray}
\chi _{1} &=&c^{\prime }(1+\lambda \eta \xi )^{z-1}, \notag \\
\chi _{2} &=&c^{\prime }r_{f}\lambda \eta (1+\lambda \eta \xi
)^{z-2}\chi
_{1}, \notag \\
&&\vdots \\
\chi _{k} &=&c^{\prime }r_{f}\lambda \eta (1+\lambda \eta \xi
)^{z-2}\chi _{k-1}, \notag
\end{eqnarray}
where the factor $r_{f}$ and the power $z-2$ are due to the fact
that there are $z-2$ local gauche conformations out of $z-1$
possible for a middle segment located anywhere except the center of
the Cayley tree, $c^{\prime }$ is a scaling factor measuring the
distance from the Cayley tree center to be evaluated in the
following. The summation in the third term of (\ref{rec GPF}) can
now be carried out to yield
\begin{equation}
\overset{x-1}{\underset{k=2}{\sum }}\chi _{k}\chi
_{x-k-2}=(x-2)(c^{\prime }r_{f}\lambda \eta )^{x-3}(1+\lambda \eta
\xi )^{(x-3)(z-2)}\chi _{1}^{2}
\end{equation}
Finally, we find
\begin{equation}
Z=c(1+\lambda \xi )^{z}+cc^{\prime
x-1}\frac{z}{2}r_{f}^{x-2}x\lambda ^{x}\eta ^{x-1}(1+\lambda \eta
\xi )^{z^{\prime }}.  \label{Bethe GPFA}
\end{equation}

Therefore, the function $\psi$ for this configuration appears to be
\begin{equation}
    c_{x}\gamma x\lambda ^{x}\eta ^{x-1}(\lambda\eta \xi_{x} )^{\theta
_{x+1}+...+\theta _{x+z^{\prime }}},
\end{equation}
where $\gamma $ is the number of ways per site the polymer chain can
be arranged on otherwise empty lattice.
We find that $\gamma _{\text{flex}}=\frac{z}{2}%
(z-1)^{x-1}$ for completely flexible chains, $\gamma
_{\text{rigid}}=\frac{z}{2}$ for rigid rods, and $\gamma
_{\text{flex}}=\frac{z}{2}[ w_{f}(z-2)+1]^{x-1}$ for semiflexible
chains. The GPF for this configuration will be
\begin{equation}
    c_{x}\gamma x\lambda^{x}\eta^{x-1}(1+\lambda \eta \xi_{x}
)^{z^{\prime }},
\end{equation}
where the constants $c_{x}$ and $\xi_{x}$ can be related to $c$ and
$\xi$, respectively: We write GPF for the aggregate with the empty
internal site and one empty surface site, and compare it with GPF
written for the aggregate with two empty sites being internal. The
former is obtained summing over all $\theta$'s on $z-1$ surface
sites, which results ${\sum}_{\{\theta_{i}\}} c(\lambda
\xi)^{\theta_{2}+...+\theta_{z}}$. The latter is evaluated by
summing over $2(z-1)$ surface sites:
\begin{equation}
    {\sum}_{\{\theta_{i},\theta_{i}^{\prime}\}} c_{2} (\lambda \xi_{2}
)^{\theta_{2} + ...+ \theta_{z} + \theta_{2}^\prime +...+
\theta_{z}^\prime}.
\end{equation}
Both partition functions should be the same,
hence the relations $\xi=\xi_{2}$ and $c=c_{2}(1+\lambda\xi)^{z-1}$
are valid, as found by performing the summation over various
configurations $\theta_{2}^{\prime},...,\theta_{z}^{\prime}$. This
idea can be applied for aggregates of larger sizes. Our suggestion:
since the chains consist of $x-1$ bonds each, the relation should be
\begin{equation}
\xi =\xi_{x}\text{ \ and \ }c=c_{x}(1+\lambda \xi )^{(z-1)(x-1)},
\label{scaling}
\end{equation}
which yields
\begin{equation}
Z=c(1+\lambda \xi )^{z}+c_{x}\gamma x\lambda ^{x}\eta
^{x-1}(1+\lambda \eta \overline{\xi })^{z^{\prime }},
\label{Behte_GPF}
\end{equation}
which is identical to (\ref{Bethe GPFA}) provided $c^{\prime
}=(1+\lambda \xi )^{-(z-1)}$.

Having established GPF, we now turn to the detailed study of the
equal probability condition for absorption at the internal site and
surface sites. Note also that this probability is the fractional
coverage of the lattice $\theta=n/N$. The following pair of
equations
\begin{eqnarray} \label{prob_cond_a}
\theta &=&Z^{-1}\underset{i_{2}i_{3}...i_{x}} {\sum}
\underset{\{{\sum} ^{j}\}} {\sum} c_{x}\lambda^{x} \eta
^{x-1}(\lambda\eta\xi_{x})^ {{\sum}^{2}\theta + ...
+{\sum}^{x}\theta}
\notag \\
&=&Z^{-1}c_{x}\gamma x\lambda ^{x}\eta ^{x-1}(1+\lambda \eta
\xi)^{z^{\prime }}
\end{eqnarray}
and
\begin{eqnarray} \label{prob_cond_b}
\theta &=&(zZ)^{-1}c \underset{\{\theta_{i}\}} {\sum }(\theta_1 +
... +\theta_z)(\lambda \xi) ^{\theta_1 +...+
\theta_z} \notag \\
&&+(zZx)^{-1}c_{x} \notag\\ 
&&\times \underset{i_{2}i_{3}...i_{x}} {\sum}
\underset{\{{\sum} ^{j}\}} {\sum} {(\underset{x-1} {\underbrace{1 +
... + 1}} + {\sum}^{2}\theta + ...+{\sum}^{x}\theta)} \notag\\
&&\times \lambda^{x}
\eta ^{x-1}(\lambda\eta\xi)^
{{\sum}^{2}\theta + ... +{\sum}^{x}\theta} \\
&=&(zZ)^{-1}c\xi \frac{\partial }{\partial \xi }(1+\lambda \xi )^{z}
+(zZx)^{-1}c_{x}\gamma \lambda ^{x}\eta ^{x-1} \notag\\
&&\times \left(2(x-1)(1+\lambda
\eta\xi)^{z^{\prime }} + \xi\frac{\partial }{\partial \xi } 
 (1+\lambda
\eta \xi )^{z^{\prime }}\right), \notag
\end{eqnarray}
are obtained in analogy with the corresponding equations from Ref.
\cite{Chang1939}. Combining Eqs. (\ref{prob_cond_a}) and
(\ref{prob_cond_b}), one finds the equations of state:
\begin{eqnarray}
\frac{\theta }{1-\theta } &=&\frac{zx}{z^{\prime }}\frac{\varepsilon
(1+\eta
\varepsilon )}{1+\varepsilon }  \label{x-mer eqsa} \\
\gamma x\eta ^{x-1}\lambda ^{x} &=&\frac{x\varepsilon }{z^{\prime
}}\frac{ (1+\varepsilon )^{x-2+z^{\prime }}}{(1+\eta \varepsilon
)^{z^{\prime }-1}}, \label{x-mer eqsb}
\end{eqnarray}
where $\varepsilon =\lambda \xi $. The solution of the quadratic
equation (\ref{x-mer eqsa}) is given by
\begin{equation}
\varepsilon (\theta )=\varepsilon _{+}=\frac{\theta z^{\prime
}-(1-\theta )zx+D}{2zx\eta (1-\theta )},  \label{eps[theta]}
\end{equation}
where
\begin{equation}
D=\sqrt{[\theta z^{\prime }-(1-\theta )zx]^{2}+4z^{\prime }zx\eta
\theta (1-\theta )}.
\end{equation}
Another solution with the negative sign before $D$ in
(\ref{eps[theta]}) is physically irrelevant, since $\varepsilon
_{-}=-1$ in the athermal limit $\eta=1$, while $\varepsilon
_{+}=z^{\prime }\theta /zx(1-\theta )>0$ in this case.

There is another method to derive (\ref{x-mer eqsa}) and (\ref{x-mer
eqsb}): The occupation $\theta$ is obtained as the ratio of the term
corresponding to the occupied internal site to the total partition
function:
\begin{equation}
\theta = Z^{-1}c\frac{z}{2}r_{f}^{x-2}x\lambda ^{x}\eta
^{x-1}(1+\lambda \xi )^{-(x-1)(z-1)}(1+\lambda \eta \xi )^{z^{\prime
}} \label{theta eq}
\end{equation}
This allows us to derive (\ref{x-mer eqsb}):
\begin{equation}
\frac{z}{2}r_{f}^{x-2}x\lambda ^{x}\eta ^{x-1}=\frac{\theta }{1-\theta }%
\frac{(1+\lambda \xi )^{x-1+z^{\prime }}}{(1+\lambda \eta \xi )^{z^{\prime }}%
},  \label{x-mer eqs bA}
\end{equation}
It should also be considered that the probability to find the
internal site of the aggregate unoccupied can be calculated by two
methods:
\begin{equation}
Z^{-1}c(1+\lambda \xi )^{z}
\end{equation}
and
\begin{eqnarray}
&&Z^{-1}c\{(1+\lambda \xi )^{z-1}+(z-1)\lambda ^{2}\eta (1+\lambda
\eta \xi
)^{z-2}\chi _{x-1} \notag \\
&&+\frac{z-2}{2}r_{f}(1+\lambda \eta \xi )^{z-3}\lambda ^{3}\eta ^{2}%
\overset{x-1}{\underset{k=2}{\sum }}\chi _{k}\chi _{x-k-1}\},
\end{eqnarray}
where the quantity in the curly brackets is the partial partition
function for the surface $z$ sites calculated under the condition
that the internal site is empty. Thus, with aid of (\ref{x-mer eqs
bA}), we obtain
\begin{equation}
(1+\lambda \xi )^{z}=(1+\lambda \xi )^{z-1}+\frac{z^{\prime }}{zx}\frac{%
\theta }{1-\theta }\frac{(1+\lambda \xi )^{z}}{(1+\lambda \eta \xi
)},
\end{equation}
which can be reduced to (\ref{x-mer eqsa}).

To this end we remark that one can also derive these equations
solely from the kinetic considerations for molecular
absorption/desorption process \cite{Roberts1937}. Here we provide
only the derivation of the first equation to save space. In the
aggregate represented by a site surrounded by $z$ sites, the
probability for a configuration $(0;\theta _{1},...,\theta _{z})$ to
occur is assumed to be proportional to $\phi ^{\theta
_{1}+...+\theta _{z}}$, where $\phi $ is the quantity to be
determined further. Then, the probability for the internal site to
be unoccupied should be proportional to ${\sum }_{\{\theta _{i}\}}
\phi ^{\theta _{1}+...+\theta _{z}}=(1+\phi )^{z}$. Hence, the
probability for one of its $z$ neighbors to be occupied is given by
\begin{equation}
\frac{\sum \theta _{1}\phi ^{\theta _{1}+...+\theta _{z}}}{(1+\phi
)^{z}}= \frac{\phi (1+\phi )^{z-1}}{(1+\phi )^{z}}=\frac{\phi
}{1+\phi }. \label{prob hole-mon}
\end{equation}
For the aggregate enclosing a chain with its neighborhood, the
probability for any of the surface sites to be occupied, provided
all internal sites are empty, is calculated in the analogous manner
and is given by
\begin{equation}
\frac{\overline{\phi }}{1+\overline{\phi }},  \label{prob hole-mon1}
\end{equation}
where $\overline{\phi }$ is the quantity similar in nature to
$\phi$. Comparing (\ref{prob hole-mon}) and (\ref{prob hole-mon1}),
one gets $\overline{\phi }=\phi $. The probabilities for various
states of occupation of the surface sites, provided that all $x$
internal sites are occupied, are proportional to ($\eta \xi
$)$^{\theta _{x+1}+...+\theta _{x+z^{\prime }}}$, which gives
\begin{equation}
\frac{\eta \overline{\phi }}{1+\eta \overline{\phi }}
\end{equation}
for the probability of a surface site to be occupied when the
internal sites are also occupied. The condition is
\begin{eqnarray}
\theta &=&(1-\theta )\frac{\phi }{1+\phi } \\
&&+\theta \left( \frac{2(x-1)}{zx}+\frac{z^{\prime }}{zx}\frac{\eta
\phi }{ 1+\eta \phi }\right) ,
\end{eqnarray}
where $[2(x-1)/zx]\theta $ is the ratio of number of bonds
$2(x-1)\theta /zx$ per lattice site to the total number of lattice
bonds $z/2$ per lattice site representing the probability to have a
bond extending from an occupied site. The equation (\ref{x-mer
eqsa}) follows immediately taking
\begin{equation}
    \phi =\varepsilon.
\end{equation}

\section{Thermodynamics}

Further analysis of (\ref{GPF}) can be carried out to derive the
configurational entropy expression. First, we write the partition as
\begin{eqnarray}
Z &=&\sum \Omega \lambda ^{n}\eta ^{X}w_{f}^{N_{g}} \\
&=&\sum e^{N(\theta g(\theta )+\theta \ln \lambda +\theta _{X}\ln
\eta+\theta _{g}\ln w_{f})},
\end{eqnarray}
where $\theta _{X}=X/N$, and $\theta _{g}=N_{g}/N$. The minimal free
energy can by achieved applying the condition:
\begin{equation}
\frac{\partial }{\partial \theta }[\theta g(\theta )+\theta \ln \lambda
+\theta _{X}\ln \eta +\theta _{g}\ln w_{f}]=0.
\end{equation}
Thus, the configurational entropy is given by
\begin{eqnarray}
S_{N}/k_{\text{B}} &=&[\theta g(\theta )]^{\ast } \notag \\
&=&-\int \ln \lambda d\theta -\theta _{X}\ln \eta -\theta _{g}\ln
w_{f}, \label{Sconf gen form}
\end{eqnarray}
where $S_{N}$ is the entropy per lattice site. Note, $g^{\ast
}(0)=0$, $\theta _{X}(0)=0$, and $\theta _{g}(0)=0$. It is
appropriate to define the entropy per monomer:
\begin{equation}
S_{n}=S_{N}/\theta .
\end{equation}
After integration of (\ref{Sconf gen form}), the entropy expression
adopts the form
\begin{eqnarray}
S_{n}/k_{\text{B}} &=&\frac{1}{x}\ln (\gamma x\eta
^{x-1})-\frac{1}{x}\ln
\theta -\frac{(1-\theta )}{\theta }\ln (1-\theta )  \notag \\
&&+\frac{(x-1)(1-\theta )}{x\theta }\ln [(1+\varepsilon )(1-\theta
)]  \notag
\\
&&+\frac{z^{\prime }}{x}\ln \theta +\frac{z^{\prime
}}{x}\frac{(1-\theta )}{
\theta }\ln (1-\theta )  \label{Sconf} \\
&&-\frac{z^{\prime }}{x}\ln \left( \frac{zx\varepsilon }{z^{\prime
}}\right) +\frac{z^{\prime }}{x}\frac{1}{2\theta }\ln \left(
\frac{1+\varepsilon }{
1-\theta }\right)  \notag \\
&&+E_{n}/k_{\text{B}}T, \notag
\end{eqnarray}
where $E_{n}/k_{\text{B}}T=-(\theta _{X}\ln \eta +\theta _{g}\ln
wf)/\theta $ is the system internal energy per monomer.

The contact density $\theta _{X}$ is evaluated making use of the
lattice topological relations:
\begin{eqnarray}
z\theta &=&2\theta _{X}+\theta _{X^{\prime }}, \\
z(1-\theta ) &=&2\theta _{X^{\prime \prime }}+\theta _{X^{\prime }},
\end{eqnarray}
where $\theta _{X^{\prime }}$ is the density of pairs formed by an
empty site adjacent to an occupied and  $\theta _{X^{\prime \prime
}}$ is the density of nearest-neighbor pairs formed by empty sites.
We calculate $\theta _{X^{\prime \prime }}$ from the fact that the
probability of having an empty site being a neighbor to another
empty site is, in accordance with (\ref{prob hole-mon}), equal to
$1/(1+\varepsilon )$, and taking into account that there are on
average $z/2$ lattice bonds per site:
\begin{equation}
\theta _{X^{\prime \prime }}=\frac{z}{2}\frac{(1-\theta )}{1+\varepsilon }.
\end{equation}
Then, we have
\begin{equation}
\theta _{X}=z\theta -\frac{z}{2}+\frac{z}{2}\frac{(1-\theta )}{1+\varepsilon
}.  \label{cont dens}
\end{equation}
The density of the gauche bonds $\theta _{g}$ is given by
\begin{equation}
\theta _{g}=\frac{\theta }{x}(x-2)\frac{(z-2)w_{f}}{1+(z-2)w_{f}},
\label{gauche dens}
\end{equation}
where $\theta /x$ is the number of polymers, $(x-2)$ is the number
of middle groups in each polymer, and the factor
\begin{equation}
f=(z-2)w_{f}/[1+(z-2)w_{f}]  \label{Flory fct}
\end{equation}
is the probability to form the gauche configuration derived by Flory
\cite{Flory1956}. The latter can be obtained from our theory as
well: The density of gauche bonds $\theta _{g}$ is the ratio of
\begin{equation}
\frac{z(z-1)}{2}-\frac{z}{2}=\frac{z(z-2)}{2}
\end{equation}
terms corresponding to the gauche configurations in (\ref{rec GPF})
to the total GPF:
\begin{eqnarray}
\theta _{g}&=&(1/Z)(z/2)(z-2)w_{f}(x-2)\lambda ^{x}\eta
^{x-1} \notag \\
&&\times(1+\lambda \xi )^{-(x-1)(z-1)}(1+\lambda \eta \xi )^{z^{\prime
}},
\end{eqnarray}
which, by use of Eq. (\ref{theta eq}), can be written in the more
compact form:
\begin{equation}
\theta _{g}=\theta \frac{x-2}{x}\frac{w_{f}}{r_{f}}.
\end{equation}
This completes our proof of (\ref{gauche dens}).

The equilibrium properties of the system can be deduced from the
thermodynamic identity
\begin{equation}
\mu =H_{n}-TS_{n},  \label{therm expr}
\end{equation}
where the configurational enthalpy per monomer,
$H_{n}=E_{n}+Pv_{0}/\theta $, with $v_{0}$ being the volume of a
lattice site. By aid of Eq. (\ref{x-mer eqsb}) the density
dependence of the chemical potential, $\mu =k_{\text{B}}T\ln \lambda
$, arises straightforwardly:
\begin{eqnarray}
\mu &=&-\frac{k_{\text{B}}T}{x}\ln (\gamma x\eta
^{x-1})+\frac{k_{\text{B}}T
}{x}\ln \left( \frac{\theta }{1-\theta }\right)  \\
&&+\frac{k_{\text{B}}T}{x}(x-1)\ln (1+\varepsilon )+z^{\prime }\ln
\left( \frac{1+\varepsilon }{1+\eta \varepsilon }\right) \notag.
\end{eqnarray}
Performing further algebraic manipulations, we derive
\begin{equation}
Pv_{0}/k_{\text{B}}T=\frac{z}{2}\ln (1+\varepsilon )+\frac{z-2}{2}\ln
(1-\theta ).  \label{P}
\end{equation}
The equation (\ref{P}) serves as the implicit equation for
determination of $\theta$.

Making use of the relations for the densities, Eq. (\ref{Sconf}) can
be written in a slightly different form
\begin{eqnarray}
S_{n}/k_{\text{B}} &=&\frac{1}{x}\ln  \frac{z}{2} + \frac{\ln
x}{x} - \frac{\ln
\theta }{x}-\frac{(1-\theta )}{\theta }\ln (1-\theta ) \notag \\
&&+\frac{(x-1)(1-\theta )}{x\theta }\ln [(1+\varepsilon )(1-\theta )] \notag \\
&&+\frac{z^{\prime }}{x}\ln \theta +\frac{z^{\prime
}}{x}\frac{(1-\theta )}{
\theta }\ln (1-\theta ) \label{Sn_final} \\
&&-\frac{z^{\prime }}{x}\ln \left( \frac{zx\varepsilon }{z^{\prime
}}\right) +\frac{z^{\prime }}{x}\frac{1}{2\theta }\ln \left(
\frac{1+\varepsilon }{
1-\theta }\right) \notag \\
&&-\frac{(\theta _{X}-\theta _{b})\ln \eta }{\theta } \notag \\
&&-\frac{x-2}{x}[ f\ln f+(1-f)\ln (1-f) \notag \\
&&-f\ln (z-2)], \notag
\end{eqnarray}
where $\theta _{b}$ is the bond density given by
$\theta_{b}=(x-1)\theta /x$.

The equations (\ref{P}) and (\ref{eps[theta]}) are to be solved
numerically for $\theta$ as a function of $T$ at constant $P$. We
then substitute $\theta $ and $\varepsilon (\theta )$ calculated
from (\ref{eps[theta]}) into (\ref{Sn_final}) to obtain the entropy
at a constant pressure. Finally, configurational heat capacity per
occupied site, $C_{p}$, can be computed from
\begin{equation}
C_{p}=(\partial H_{n}/\partial T)_{p}
\end{equation}
or equally from
\begin{equation}
C_{p}=T(\partial S_{n}/\partial T)_{p}.
\end{equation}
According to our numerical computation, both equations give
identical specific heat values.

\section{Comparison with Other Results}

In the present section, we analyze the main characteristics of the
thermodynamic functions (\ref{P}) and (\ref{Sn_final}) in comparison
with data available from experiments and simulations.

\begin{figure}
\epsfxsize=3.4in \epsffile{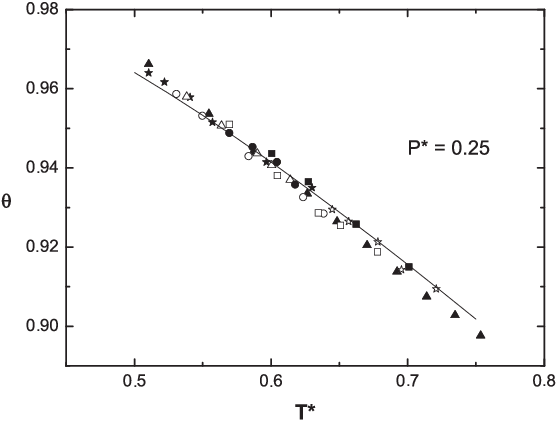}%
\caption{Density $\theta$ as a function of reduced temperature $T^{*}$ for constant reduced pressure $P^{*}=0.25$. Experimental data for various
polymer liquids are from \cite{sanchezlacombe3}, the line is the solution of Eq. (\ref{P}) for $z=20$.} \label{SanchezLacome}
\end{figure}

Presented as early as in 1976, the lattice fluid (LF) theory
developed by Sanchez and Lacombe \cite{sanchezlacombe1,sanchezlacombe1a}, due to its
simplicity and high level of prediction, became a foundation of many
successional theoretical and experimental investigations of polymer
systems \cite{sanchezlacombe2}. In analogy with LF theory, the
equation of state (\ref{P}) can be written in terms of the reduced
temperature $T^{*}=-2k_{B}T/zV$ and pressure $P^{*}=-2v_{0}P/zV$.
Figure \ref{SanchezLacome} shows the experimental data for density,
$\theta$, vs. reduced temperature, $T^{*}$, obtained from in Fig. 3
of Ref. \cite{sanchezlacombe3} for various liquids. The line is the
theoretical isobar ($P^{*}=0.25$) computed applying the limit
$x\rightarrow\infty$ for $z=20$.
Based on this we conclude that the present theory adequately
describes the relationship of density, temperature, and pressure for
pure polymer fluid.

The important difference between the theory by Sanchez and Lacombe
and the present theory is in entropy. The LF entropy is independent
of temperature, while (\ref{Sn_final}) exhibits a temperature
dependence. This is a consequence of consideration of the
configurational distribution of the polymer system influenced by
mer-mer interaction energy and polymer semiflexibility, which is
absent in LF theory.

\begin{figure}
\epsfxsize=3.4in \epsffile{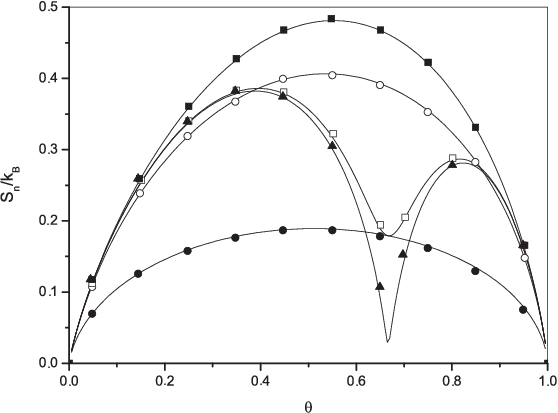}%
\caption{Configurational entropy of dimers on one-dimensional lattice at various
interaction energies. The curves are predictions of Bethe-Peierls theory with $z=2$. The data points are from \cite{Davila}: filled circles, open circles, filled squares, open squares, filled triangles - $V=-5,-2,0,2,5$, respectively.} \label{FigSvsTheta}
\end{figure}

Recently, Davila \emph{et al.}\cite{Davila} presented Monte Carlo (MC) data for
interacting dimers occupying one-dimensional, square, triangular and hexagonal lattice. Our configurational entropy
expression (\ref{Sn_final}), with an appropriate integer value for $z$,
is in perfect agreement with the quasi-chemical approximation.
In particular, Bethe-Peierls approximation becomes exact in
the case of dimers on the one-dimensional lattice, see Fig. \ref{FigSvsTheta}.
Essentially, this is a numerical verification of the fact known to Chang
\cite{Chang1939}. The test of the present theory against the MC data shows a significant improvement over the Brag-Williams approximation. In addition, it shows some special features such as correspondence between $z$ and the coordination number of the regular lattice, i.e. $z=2,3,4,6$ for one-dimensional, honeycomb, square and triangular lattice, respectively. Chemical and physical topology  of the adsorbing surface is known\cite{sasha1,sasha2} to be critically important for the adsorption ability of the surface. This local surface inhomogeneity may be attributed to an inhomogeneity of $z$ and could be treated using the recursive lattice approach as well \cite{Chhajer1,Chhajer2,Chhajer3}.

\section{Specific Heat}
One of applications of this theory could be a comparison with
experimental results available from literature for specific heat
dependence on molecular weight and temperature. The specific heat of
linear \emph{n}-alkanes as a function of the number of carbons in
the backbone is found to be a monotonically decreasing function
\cite{McKenna1,McKenna2}, while the specific heat of a continuously
polymerizing system was measured to pass through a maximum
\cite{Johari}. We suggest that the controversy can be resolved in
the framework of the present theory by studying the configurational
specific heat as a function of chain length.
\begin{figure}
\epsfxsize=3.4in \epsffile{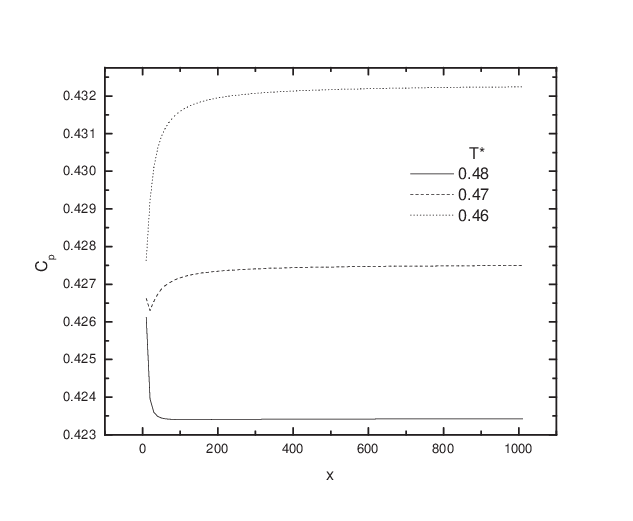}%
\caption{Three typical dependencies of the configurational specific
heat on chain length. For this figure we used $T^{*}_{f}=0.5$,
$P^{*}=0.25$, $z=12$.} \label{FigCpvsX}
\end{figure}
Figure \ref{FigCpvsX} shows $C_{p}$ vs. $x$ in three cases:
$T^{*}=0.48,0.47,0.46$ represented by continuous, dashed, and dotted
lines, respectively; in all three cases $T^{*}_{f} = 0.5$, where $T^{*}_{f}\equiv zU_{f}/2$. By showing
this we want to demonstrate three types of typical behavior: for
high temperatures ($T^{*}=0.48$ and above) the curve is always
sharply decreasing, then reaching a constant value; for low
temperatures ($T^{*}=0.46$ and below) the curve is monotonically
increasing; in the intermediate region, the curve initially
decreases reaches a minimum and then increases. As was shown by
Di Marzio and Dowell \cite{DiMarzioDowell}, the
specific heat has a large vibrational part, hence the above experimental
results cannot directly be fitted by the theoretical curves at current stage. However, Johari \emph{et. al.}\cite{VibrSpecJohariEtAl1,VibrSpecJohariEtAl2,VibrSpecJohariEtAl3}
found experimentally that the vibrational specific heat of glass and liquid are practically the same.
Accordingly, the configurational specific heat is useful for estimating the
jump in heat capacity at the glass transition, provided the specific heat of glass is measured.

\section{Discussion and Conclusion}
We provide a very careful derivation of the entropy and the pressure
expressions based on the Bethe-Peierls approximation for linear
monodisperse polymer. The prediction of the system density at a
given reduced pressure and temperature is shown to be in
quantitative agreement the experimental data and with Monte Carlo
simulations on square lattice. The overall configurational
entropy of the polymer fluid consists of three fundamentally
different types of contributions, i.e., the athermal entropy
(translational and configurational degrees of freedom), the
semiflexibility correction entropy, and the thermal correction
entropy that is due to the thermal correction term of the partition
function. The third type, that is absent in Flory-Huggins theory and
LF theory by Sanchez and Lacombe, contributes the main part at low
temperatures. We also demonstrate that our theory is equivalent to
the Ryu-Gujrati theory with the correction due to chain
semiflexibility.

We obtain the negative entropy states at certain conditions: low temperatures,
$T\subset[0..T_{\text{K}}]$, and some fixed pressure; high
pressures, $P\subset[P_{\text{K}}...\infty)$, and a fixed
temperature; high molecular weights,
$x\subset[x_{\text{K}}..\infty)$, while both temperature and
pressure are kept constant. These results are in qualitative
agreement with the earlier work by Di Marzio \emph{et al.}
\cite{Gibbs-DiMarzio,DiMarzioEtAl}. The entropy crisis was captured at low
temperatures comparable to energy due to polymer semiflexibility. This
indicates that the local anisotropy due to semiflexibility could the rout
cause of entropy crisis in our calculation.

Our computation of the configurational specific heat as a function
of chain length revealed that this function is monotonically
increasing at low temperatures and monotonically decreasing at high
temperatures. In the transition region the function has a minimum.

Recently, it was demonstrated \cite{Schroeder-Turketal}
that the local anisotropy plays an important role for the jamming effect in spherical
bead systems. This research is in accordance with our ideas since we are not
concerned with shapes of particles or local occupation probabilities,
but rather consider polymer semiflexibility. However, it is not completely clear how the
dynamic arrest is caused by the local anisotropy. To address this problem we mention
the result in \cite{Semeriyanovetal}, where it was shown that the
absence of lattice loops may amplify the effect of local anisotropy on averaged
transport properties, which is shown for percolating resistors randomly distributed on the lattice.
The breakage of the local anisotropy at the glass transition for the spherical beads can be
explained in the framework of the theory of glass transition of Edwards and Vilgis
\cite{EdwardsVilgis}, where the dynamic arrest is caused by closed path configurations.
The loops get frozen first. This makes the transport of excitations
preferentially along tree-like structures. Hence, the system creates more anisotropic local configurations due to their higher mobility in the tree
environment. The theory of Edwards and Vilgis shows how the dynamical properties of continuous models
and equations can be transferred into discrete counterparts. Thus, we think that the
configurational entropy, obtained by a proper discretization scheme, can be used to
predict the rapid slowing down. Also, we suggest that the Nernst postulate formulated for the
isotropic liquids should be supplemented by the idea of local anisotropy.

\section{Acknowledgments}
FFS is grateful for hospitality and support to Leibniz Institute
of Polymer Research Dresden.

\section{Appendix}

In this appendix we demonstrate that the entropy expression
developed in the paper by Ryu and Gujrati \cite{Ryu-Gujrati1997}
with added correction due to polymer semiflexibility is equivalent
to (\ref{Sn_final}). Formulated for multicomponent mixtures their
theory can easily be adopted to our system with minor corrections.
One of these is due to discrepancy in methodology of accounting for
the mer-mer interaction. In their theory this interaction is
incorporated (as in the Flory-Huggins theory) through the
\emph{effective} interaction between occupied sites and vacancies
with energy given by
\begin{equation*}
    \epsilon_{10} = e_{10}-1/2(e_{11}+e_{00}),
\end{equation*}
where $e_{10},e_{11},e_{00}$ are the energies of physical
interactions for mer-hole, mer-mer and hole-hole pairs,
respectively. Since only the second is actually present,
$e_{01}=0,e_{11}=V,e_{00}=0$, one gets $\epsilon_{10} = -V/2$, with
$w=\exp(V/2RT)$ being the corresponding Boltzmann weight. Note also
that $q$ in their symbolic representation is equivalent to $z$ in
ours both signifying the coordination number. The semiflexibility
can be incorporated by replacement of $r=q-1$ in their paper by
$r_{f}$ defined in (\ref{r_f}) and by adding the term $-N_{g}\ln
w_{f}$ to the entropy [see Eq. (\ref{S_n_RG}) below]. Here we
summarize the results of their theory. The final equations resulting
from the Bethe lattice iterative technique are the following:
\begin{eqnarray*}
y_{0,1}&=&(w + \lambda_{1} y_{1}^{x-1} y_{0,1}^{z-1} / r_{f}) / \\
&& \> \> \> \>(1
+w \lambda_{1} y_{1}^{x-1} y_{0,1}^{z-1} / r_{f}), \\
1/(1-\theta)&=&1+(z/2) x y_{0,1}^{z} y_{1}^{x-1} /
r_{f},\\
Q_{1} &=& 1 + w \lambda_{1} y_{1}^{x-1} y_{0,1}^{z-1} / r_{f},
\end{eqnarray*}
where $\lambda_{1}=(z/2)x-(x-1)$ and the parameters
$y_{1},y_{0,1},Q_{1}$ are evaluated from the relations
\begin{eqnarray*}
y_{0,1} &=& \left[ -w(p-1) + \sqrt{w^{2}(p-1)^2+4p} \right]/2, \\
Q_{1} &=& 1+wp/y_{0,1},
\end{eqnarray*}
with $p=\theta\lambda_{1}/(z/2)(1-\theta)x$. Below we list the
expressions for the the densities which are numbers of the
corresponding species per lattice site:
\begin{gather*}
    \phi_{n,1}= \theta / x,\> \> \> \>\phi_{b} = \phi_{n,1} (x-1),\>
    \> \> \>
\phi_{00} = z(1-\theta)/Q1,
\\ \phi_{01} = z(1-\theta)-2\phi_{00}, \> \> \> \> 
    \phi_{11} = (z/2)\theta - \phi_{01} - \phi_{b},
\\ \phi_{1u} =
(z/2)\theta - \phi_{b},\> \> \> \> \phi_{0u} = (z/2)(1-\theta), 
\> \> \> \>  \phi_{u}
= z/2 - \phi_{b},
\\ \phi_{11}^{0} =
\phi_{1u}^{2}/\phi_{u},\> \> \> \> \phi_{00}^{0} =
\phi_{0u}^{2}/\phi_{u},\> \> \> \> \phi_{01}^{0} =
2\phi_{0u}\phi_{1u}/\phi_{u},
\end{gather*}
where $\phi_{n,1}$ and $\phi_{b}$ are the densities of polymers and
bonds, respectively; $\phi_{00}$, $\phi_{01}$ and $\phi_{11}$ are
the densities of hole-hole, hole-mer and mer-mer nearest-neighbor
contacts; $\phi_{00}^{0}$, $\phi_{01}^{0}$ and $\phi_{11}^{0}$ are
the corresponding athermal contact densities; $\phi_{1u}$ and
$\phi_{0u}$ are the densities of unoccupied lattice bonds associated
with occupied sites and vacancies, respectively; $\phi_{u}$ is the
density of unoccupied lattice bonds. The adimentional pressure,
$\omega=\beta Pv_{0}$ ($\beta=1/k_{\text{B}}T$), given by
\begin{equation*}
\beta Pv_{0} = -\ln (1-\theta) + (z/2)\ln(2\phi_{u} / z) + (z/2)
\ln(\phi_{00}^{0}/\phi_{00}).
\end{equation*}
can be transformed to the form equivalent to (\ref{P}):
\begin{equation}
\beta Pv_{0} = -\ln(1-\theta)+(z/2)\ln [(1-\theta)Q_{1}].
\label{omega_RG}
\end{equation}
Finally, we compared the Ryu-Gujrati configurational entropy
corrected to account for semiflexibility
\begin{eqnarray}
S_{n}/k_{\text{B}} &=& ( \phi_{n,1} \ln[(z/2) r_{f}^{x-2} x /
\phi_{n,1}] - (1-\theta)\ln(1-\theta) \notag \\
&+& \phi_{01} \ln(\phi_{01}^{0} / \phi_{01}) + \phi_{00}
\ln(\phi_{00}^{0} / \phi_{00}) \label{S_n_RG} \\
&+& \phi_{11} \ln(\phi_{11}^{0} 
\phi_{11}) 
- N_g \ln w_{f}) / \theta  \notag
\end{eqnarray}
with (\ref{Sn_final}) and verified numerically their equivalency.

\end{document}